\documentclass[useAMS,usenatbib]{mn2e}
\usepackage{graphicx}
\usepackage{verbatim}
\usepackage{fix2col}
\usepackage{amssymb}

\title[Optical properties of SDSS galaxy classes]{The nature of the SDSS galaxies in various classes based on morphology, colour and spectral features -- I. Optical properties}
\author[J. H. Lee et al.]{Joon Hyeop Lee$^{1,2}$\thanks{E-mail:
jhl@kasi.re.kr (JHL); mglee@astrog.snu.ac.kr (MGL); cbp@kias.re.kr (CBP); yychoi@kias.re.kr (YYC)}, Myung Gyoon Lee$^{1\star}$, Changbom Park$^{3\star}$, Yun-Young Choi$^{4\star}$\\
$^{1}$Astronomy Program, Department of Physics and Astronomy, Seoul National University, Seoul 151-742, Korea\\
$^{2}$Korea Astronomy and Space Science Institute, Daejeon 305-348, Korea\\
$^{3}$Korea Institute for Advanced Study, Dongdaemun-gu, Seoul 103-722, Korea\\
$^{4}$Astrophysical Research Centre for the Structure and Evolution of the Cosmos, Sejong University, Seoul 143-747, Korea}
\begin{document}

\date{Accepted 2008 June 28. Received 2008 June 04; in original form 2008 April 14}

\pagerange{\pageref{firstpage}--\pageref{lastpage}} \pubyear{2008}

\maketitle

\label{firstpage}

\begin{abstract}
We present a comprehensive study of the nature of the SDSS galaxies 
divided into various classes based on their morphology, colour, and spectral features.
The SDSS galaxies are classified into early-type and late-type; red and blue;
passive, H{\protect\scriptsize II}, Seyfert, and LINER, returning a total of 16 fine classes of galaxies.
We examine the luminosity dependence of seven physical parameters of galaxies 
in each class.
We find that more than half of red early-type galaxies (REGs) have star formation 
or AGN activity, and that these active REGs have smaller axis ratio and bluer
outside compared to the passive REGs. Blue early-type galaxies (BEGs) show
structural features similar to those of REGs, but their centres are bluer
than REGs. H{\protect\scriptsize II} BEGs are found to have bluer centres than passive BEGs, but
H{\protect\scriptsize II} REGs have bluer outside than passive REGs.
Bulge-dominated late-type galaxies have red colours. Passive red late-types are 
similar to REGs in several aspects. Most blue late-type galaxies (BLGs) have 
forming stars, but a small fraction of BLGs do not show evidence for current
star formation activity.  Differences of other physical parameters among different classes are inspected, and their implication on galaxy evolution is discussed.
\end{abstract}


\section{Introduction}

One of the fundamental issues of the observational cosmology is the evolutionary connection between various classes of galaxies.
Today, galaxies are classified with various criteria.
The most classical classification of galaxies is the Hubble sequence: elliptical galaxies, lenticular galaxies, spiral galaxies, barred spiral galaxies, and irregular galaxies \citep{hub36,san61,dev74}.
The main criterion of this classification is the morphology of galaxies; that is, the existence, size ratio, and appearance of spiral arms, disc, bulge, and bar.
The Hubble sequence was established mainly based on the galaxies without nuclear activity, and such galaxies are often called \emph{normal} galaxies. On the other hand, more and more galaxies showing active nuclei with broad line emission are being found; they are often called \emph{abnormal} galaxies.
According to the features of the line emission, active galactic nuclei (AGN) host galaxies are classified into several sub-classes: Seyfert 1 galaxies, Seyfert 2 galaxies, broad-line radio galaxies, narrow-line radio galaxies, low ionisation nuclear emission regions (LINERs), and so on.
Luminosity is another criterion to classify galaxies. \citet{san84} defined the galaxies with M$_{B} > -18$ as \emph{dwarf} galaxies.
In addition to those general classes of galaxies, there are some unusual galaxy classes with interesting properties: E+A galaxies with the spectral features of both very old stellar populations and very young stellar populations \citep{dre92,yan06,yan08}; ultra-luminous infrared galaxies (ULIRGs) that are very bright in the mid- and far-infrared bands \citep{san96,hwa07}; blue compact galaxies that show very compact morphology and high surface brightness with blue colour \citep{thu97}; extremely red objects (EROs) whose optical $-$ infrared colour is extremely red \citep{els88}; and so on.
Recently, to deal with a large amount of survey data, many astronomers have classified galaxies simply into red sequence galaxies and blue sequence galaxies according to their colour-magnitude relation \citep[e.g.][]{mar07}.

For a long time,  the individual properties of galaxies in those various classes have been investigated, and many efforts have been made to answer fundamental questions about galaxy evolution and connections between galaxy classes. Some examples of such fundamental questions are as follows: Why is there a conspicuous bimodality in the colour distribution of galaxies? Why does the colour bimodality not exactly agree with the morphological segregation? How do the environments affect the properties of galaxies? How did active galactic nuclei (AGNs) form? Is there any transition among the different classes of galaxies?
Recent studies using large survey data discovered several interesting aspects of galaxy evolution, providing some answers to those fundamental questions.

For example, \citet{bal06} studied the bivariate luminosity functions with galaxy type classification, and found that there is a clear morphological bimodality supporting the idea that merger and accretion are associated with bulges and discs, respectively. \citet{ber05} showed that both luminosity and colour of early-type galaxies are correlated with stellar velocity dispersion, and that velocity dispersion may be also closely correlated with the age and metal abundance of early-type galaxies.
\citet{cho07} found that late-type galaxies show wider dispersion in several physical quantities than early-type galaxies, and that those physical quantities manifest different behaviours across $M_{\star}\pm1$ mag.
\citet{par07} investigated the environmental effects on various physical properties of galaxies, finding that a key constraint on galaxy formation models is the morphology-density-luminosity relation.
\citet{mat06} analysed the colour, 4000{\AA} break, and age in various spectral classes of galaxies, suggesting that the median light-weighted stellar age of galaxies is directly responsible for the colour bimodality in the galaxy population.

Those previous studies, however, are not yet enough to explain the origins of the various galaxy classes and the evolutionary connections between the classes.
One of the major limitations is that the galaxy classifications in most previous studies were often limited to only one or two properties.
For instance, automatically-classified morphology \citep[e.g.][]{par05,bal06}, galaxy colour \citep[e.g.][]{mar07}, or spectral line features \citep[e.g.][]{mat06} are used frequently in recent studies based on large survey data, but a multilateral classification study using all these criteria at the same time has not yet been seen.
Such simplifications may be often very useful to understand the various and complicated phenomena in galaxy evolution. However, a simple classification can not distinguish detailed aspects of galaxy evolution. For example, early-type galaxies are often regarded as red and passive galaxies in most studies, but that is not always true. Some kinds of galaxies with early-type morphology were found to have blue colours possibly originating from young stellar populations or active nuclei \citep{abr99,im01,fer05,lee06}. It is difficult to understand these kinds of unusual classes in studies with simple classifications.

Closely related to galaxy classification, a couple of studies searching for principal components of galaxy properties have been conducted recently.
\citet{ell05} examined the distribution of photometric, spectroscopic and structural parameters for 350 nearby galaxies using the Millennium Galaxy Catalogue \citep{lis03}, arguing that most properties show a clear distinction between early-type galaxies and late-type galaxies.
Later, \citet{con06} carried out principal-component analyses of the properties of 22000 galaxies at $z<0.05$ using the Third Reference Catalogue of Bright Galaxies \citep[RC3;][]{dev91}, finding that the three parameters determining a galaxy's physical state may be mass, star formation and interactions/mergers.
These studies presented how useful to consider various parameters in galaxy classification, although some important components like AGNs were not considered in their analyses.
\citet{con06} showed that multiple independent components, rather than any single component, may determine the various properties of galaxies.

Therefore, to understand the nature of galaxies more comprehensively and to give stronger constraints on galaxy evolution models to \emph{explain all kinds of galaxies}, it is necessary to investigate the nature of galaxies in various and finely-divided classes, and to find out the evolutionary connections between the classes.
We have been doing a comprehensive study on a set of fine galaxy classes in the Sloan Digital Sky Survey \citep[SDSS;][]{yor00}, based on their morphology, colour and spectral features.
In this paper, the first in the series, we present the optical properties of galaxies in various fine classes.
The outline of this paper is as follows. 
Section 2 shows the data set we used, and \S3 describes the methods to classify the SDSS galaxies and to select volume-limited samples.
We present the statistics of selected optical parameters and their luminosity dependence in \S4. Based on the optical properties, we discuss the nature of galaxies in the fine classes in \S5. Finally, the conclusions in this paper are given in \S6.
Throughout this paper, we adopt the cosmological parameters 
$h=0.7$, $\Omega_{\Lambda}=0.7$, and $\Omega_{M}=0.3$.

\section{Data and physical parameters}

We use the SDSS Data Release 4 \citep[DR4;][]{ade06}\footnote{See http://www.sdss.org/dr4/.} in this study. The DR4 imaging data cover about 6670 deg$^{2}$ in the $ugriz$ bands, and the DR4 spectroscopic data cover 4783 deg$^{2}$. The photometric and spectroscopic observations were conducted with the 2.5-m SDSS telescope at the Apache Point Observatory in New Mexico, USA. The median width of the point-spread-function in the $r$ band photometry is $1.4''$, and the wavelength coverage in the spectroscopy is $3800 - 9200${\AA}.
We use the photometric and structural parameters from the SDSS pipeline \citep{sto02} data, and the spectroscopic parameters from the Max-Planck-Institute for Astronomy catalogue \citep[MPA catalogue;][]{kau03,tre04,gal06}.
In addition, we use the velocity dispersion estimated using an automated spectroscopic
pipeline called idlspec2d version 5 (D. Schlegel et al. 2008, in preparation), and colour gradient, inverse concentration, and axis ratio of the SDSS DR4plus \citep{cho07} sample, which is one of the products of the New York University Value-Added Galaxy Catalogue \citep{bla05}.

After foreground extinction correction \citep{sch98},
the magnitude of each galaxy was corrected in two more aspects: the redshift effect (k-correction) and galaxy luminosity evolution (evolutionary correction). We used the method of \citet{bla03} to conduct k-correction, and the empirical formula of \citet{teg04} to conduct evolutionary correction. Using these methods, we corrected the observed magnitude of each galaxy into the magnitude at redshift z = 0.1, where the SDSS galaxies were observed most frequently.
Since the corrections are applied optionally in this paper, we denote the magnitude with k-correction as $^{0.1{\textrm{\protect\scriptsize K}}}m$ and the magnitude with both k-correction and evolutionary correction as $^{0.1{\textrm{\protect\scriptsize KE}}}m$, if the observed magnitude is $m$. We use Petrosian magnitudes to represent optical brightness, while we use model magnitudes to calculate galaxy colours.

The colour gradient $\Delta ^{0.1{\textrm{\protect\scriptsize K}}}(g-i)$ is defined as the difference in $(g-i)$ colour between the region at $0.5R_{\textrm{\protect\scriptsize pet}}<R<R_{\textrm{\protect\scriptsize pet}}$ and the region at $R<0.5R_{\textrm{\protect\scriptsize pet}}$  \citep[negative $\Delta (g-i)$ for blue outside;][]{cho07}.
The inverse concentration (R50/R90) and axis ratio of galaxies were estimated using ellipsoidal fitting \citep{cho07}.\footnote{R$_{pet}$ is the Petrosian radius, and R50 (R90) is the semimajor axis length of an ellipse containing 50 per cent (90 per cent) of the Petrosian flux.} The colour gradient, concentration, and axis ratio are corrected for seeing effects. The equivalent width of H${\protect\scriptsize \alpha}$ emission and the 4000{\AA} break index, D$_{n}$(4000), defined in \citet{bru83}, were retrieved from the MPA catalogue.

\section{Analysis}

\subsection{Galaxy classification}\label{sclass}

In this study, we classified galaxies with three criteria: morphology, colour, and spectral line features.

\begin{figure}
\includegraphics[width=84mm]{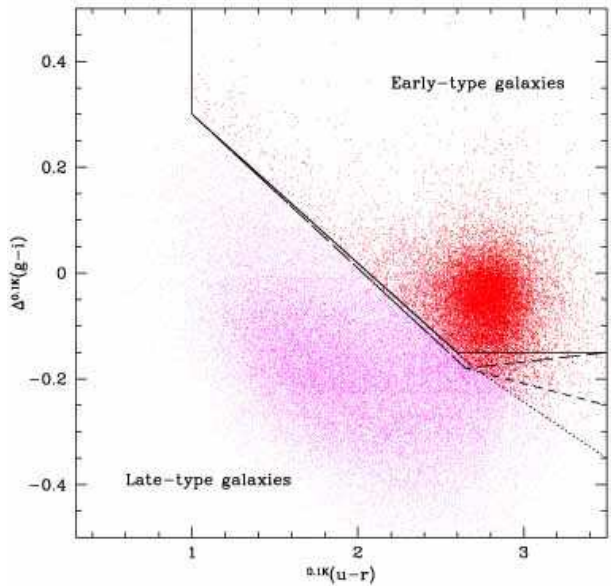}
\caption{Segregation between early-type galaxies (dark points) and late-type galaxies (light points)
in the colour vs. colour-gradient space. 
The lines represent different segregation guidelines for different magnitude ranges
(solid line for $14.5<r_{\textrm{\protect\scriptsize pet}}<16.0$, short-dashed line for $16.0<r_{\textrm{\protect\scriptsize pet}}<16.5$,
long-dashed line for $16.5<r_{\textrm{\protect\scriptsize pet}}<17.0$, and dotted line for $17.0<r_{\textrm{\protect\scriptsize pet}}<17.5$).}
\label{class1}
\end{figure}

Morphology is one of the most fundamental criteria to classify galaxies.
We selected early-type galaxies with the galaxy classification method using the colour versus colour-gradient space \citep{par05} as shown in Fig. \ref{class1}. 
In this classification method, colours and colour gradients are the main criteria for classification, and the inverse concentration cut is also applied differentially for different magnitude ranges: R50/R90 $<0.43$ for $14.5<r_{\textrm{\protect\scriptsize pet}}<16.0$, R50/R90 $<0.45$ for $16.0<r_{\textrm{\protect\scriptsize pet}}<16.5$, R50/R90 $<0.47$ for $16.5<r_{\textrm{\protect\scriptsize pet}}<17.0$, and R50/R90 $<0.48$ for $17.0<r_{\textrm{\protect\scriptsize pet}}<17.5$.

\begin{figure}
\includegraphics[width=84mm]{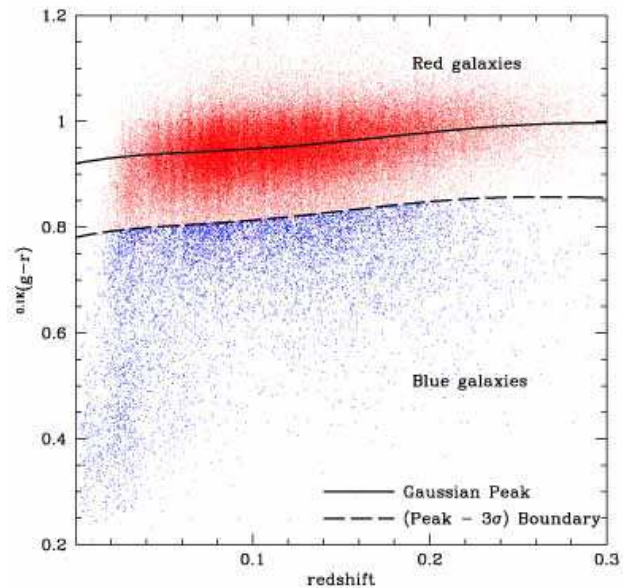}
\caption{The $^{0.1\textrm{\protect\scriptsize K}}(g-r)$ colour versus redshift of early-type galaxies. 
The solid line is the polynomial fit of Gaussian peaks in the $^{0.1\textrm{\protect\scriptsize K}}(g-r)$ distribution of early-type galaxies at 8 redshift bins between $z=0$ and $z=0.4$.
The dashed line is the (peak $- 3\sigma$) line, which is used for red/blue galaxy separation.}
\label{class2}
\end{figure}

The galaxy colour is another classification criterion that is frequently used in large galaxy surveys.
We classified the galaxies into red galaxies and blue galaxies, based on the ($g-r$) colour. Since the $r$ band is the standard band in the SDSS photometry and the photometric uncertainty in the $u$ band is relatively large compared to that in the $g$ band, we selected the ($g-r$) colour as the index for the galaxy colour segregation. We used the method of \citet{lee06} to segregate blue galaxies from red galaxies.
First, we divided the redshift range of $0 \le z \le 0.4$ into eight bins, and derived a $^{0.1\textrm{\protect\scriptsize K}}(g-r)$ colour histogram of the early-type galaxies with $^{0.1\textrm{\protect\scriptsize K}}M_{\textrm{\protect\scriptsize pet}}(r)\le-20$ for each redshift bin.
Then, we fitted the colour distribution with a Gaussian function in each redshift bin, and derived
a guideline for colour separation by fitting a 5th-order polynomial function to the colours corresponding to the peaks in the eight bins, as shown in Fig. \ref{class2}. We used a polynomial function instead of a linear function for more accurate fitting, although the resulting peak line seems almost linear at $z<0.3$. It is noted that the peak line shows a somewhat complex curve when it extends to higher redshift \citep[$z\sim1$;][]{lee06}.
We regard the early-type galaxies that are $3 \sigma$ bluer than the Gaussian peak colour as \emph{blue} early-type galaxies, and the other early-type galaxies as \emph{red} early-type galaxies.
The same colour guideline was used to separate the \emph{red} late-type galaxies and the \emph{blue} late-type galaxies.

\begin{figure}
\includegraphics[width=84mm]{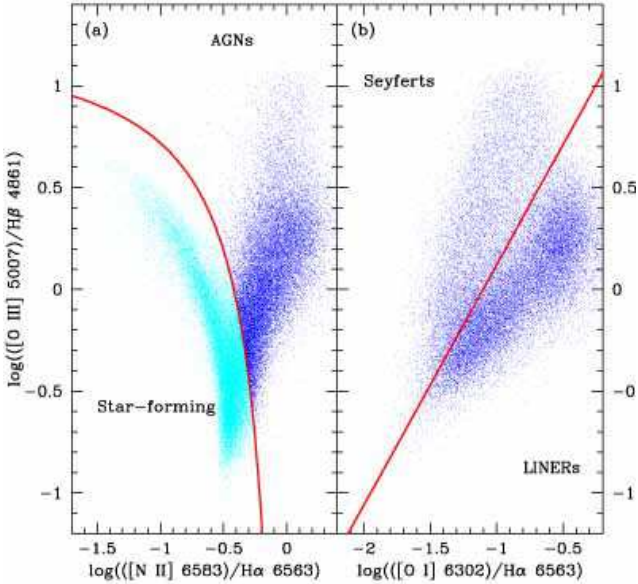}
\caption{(a) AGN selection in the BPT diagram \citep{bal81}. 
We used an empirical criterion \citep[{\it solid line}; given by][]{kau03b} to distinguish AGNs from star-forming galaxies.
(b) Seyfert-LINER segregation in the [O{\protect\scriptsize III}]/H{\protect\scriptsize $\beta$} vs. [O{\protect\scriptsize I}]/H{\protect\scriptsize $\alpha$} diagram.
The solid line is the criterion to distinguish between Seyferts and LINERs, given by \citet{kew06}.}
\label{class3}
\end{figure}

In spectroscopic studies, the line features of galaxies are often used to classify galaxies.
Based on the spectral line features, we classified the galaxies into passive galaxies, H{\protect\scriptsize II} galaxies, Seyfert galaxies and LINER galaxies.

First, We selected AGN host galaxies using the line flux ratio diagram of [O{\protect\scriptsize III}]/H${\protect\scriptsize \beta}$ versus [N{\protect\scriptsize II}]/H${\protect\scriptsize \alpha}$ \citep[BPT diagram;][]{bal81} as shown in Fig. \ref{class3}a. We used an empirical criterion to segregate AGN host galaxies from star-forming galaxies:
[O{\protect\scriptsize III}]$/$H${\protect\scriptsize \beta}$ $=0.61/$([N{\protect\scriptsize II}]$/$H${\protect\scriptsize \alpha}$ $-0.05)+1.3$, given by \citet{kau03b}. AGN selection was conducted for the sample of galaxies with a signal-to-noise ratio (S/N) of $\ge3$ for H${\protect\scriptsize \alpha}$, H${\protect\scriptsize \beta}$, [O{\protect\scriptsize III}] and [N{\protect\scriptsize II}] lines \citep{kau03b}, but we classified some galaxies as AGNs even if their S/N ratios of one or two lines are smaller than 3, in some special cases. For example, a galaxy with log([N{\protect\scriptsize II}]/H${\protect\scriptsize \alpha}$) $>-0.2$ and its S/N ratios of [N{\protect\scriptsize II}] and H${\protect\scriptsize \alpha}$ larger than 3 but with its S/N ratios of [O{\protect\scriptsize III}] and H${\protect\scriptsize \beta}$ smaller than 3, was classified as an AGN host galaxy. These S/N criteria are too generous if we intend to select a genuine sample of AGNs, but they may be useful to reduce the contamination of H{\protect\scriptsize II} galaxies due to AGNs.

Second, among the selected AGN host galaxies, we distinguished Seyferts from LINERs in the [O{\protect\scriptsize III}]/H${\protect\scriptsize \beta}$ versus [O{\protect\scriptsize I}]/H${\protect\scriptsize \alpha}$ diagram. We used an empirical guideline: [O{\protect\scriptsize III}]/H${\protect\scriptsize \beta}$ $=1.18$ [O{\protect\scriptsize I}]/H${\protect\scriptsize \alpha}$ $+1.3$, given by \citet{kew06}. AGN host galaxies in the LINER domain were classified as LINER galaxies, and the AGN host galaxies that are not LINER galaxies were classified as Seyfert galaxies.
In other words, we classified AGN host galaxies without the signal of [O{\protect\scriptsize I}] emission line as Seyfert galaxies in this paper.

Third, we selected as H{\protect\scriptsize II} galaxies, non-AGN galaxies with H${\protect\scriptsize \alpha}$ emission with S/N $\ge3$. This criterion is more generous than those in previous studies, but useful to reduce the contamination of passive galaxies due to H{\protect\scriptsize II} galaxies. Finally, passive galaxies were selected as galaxies with no or insufficient (S/N$<3$) signal of H${\protect\scriptsize \alpha}$ emission.

\begin{table}
\centering
\caption{Abbreviations of the 16 fine galaxy classes}
\label{tclcl}
\begin{tabular}{lcccc}
\hline \hline
& \multicolumn{2}{c}{Early-type} & \multicolumn{2}{c}{Late-type} \\
& Red & Blue & Red & Blue \\
\hline
Passive & $p$REG & $p$BEG & $p$RLG & $p$BLG \\
H{\protect\scriptsize II} & $h$REG & $h$BEG & $h$RLG & $h$BLG \\
Seyfert & $s$REG & $s$BEG & $s$RLG & $s$BLG \\
LINER & $l$REG & $l$BEG & $l$RLG & $l$BLG \\
\hline \hline
\end{tabular}
\end{table}

\begin{figure*}
\includegraphics[width=168mm]{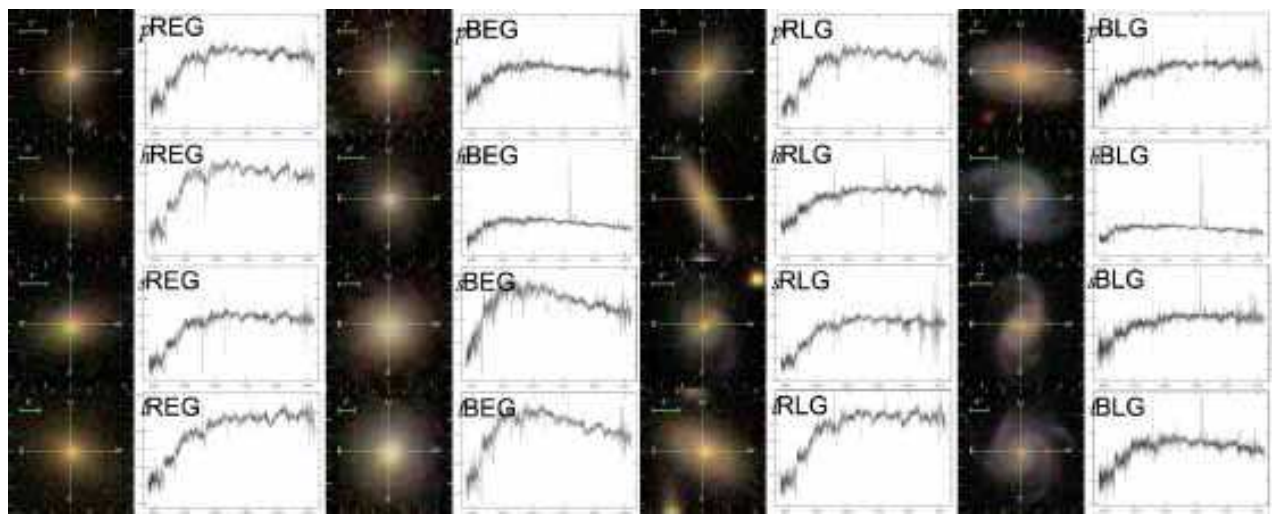}
\caption{Atlas images and spectra of sample galaxies in the 16 fine classes. These images were retrieved from the SDSS. }
\label{atlas}
\end{figure*}

Based on the three independent classifications of galaxies, we classified the galaxies into the final 16 classes: [early-type, late-type] $\times$ [red, blue] $\times$ [passive, H{\protect\scriptsize II}, Seyfert, LINER].
Hereafter, we use the following abbreviations for the 16 galaxy classes: REG (red early-type galaxy), BEG (blue early-type galaxy), RLG (red late-type galaxy), BLG (blue late-type galaxy), and $p$- (passive), $h$- (H{\protect\scriptsize II}), $s$- (Seyfert), $l$- (LINER), as shown in Table \ref{tclcl}. For example, `$p$REGs' represents `passive red early-type galaxies' and `$s$BLGs' represents `Seyfert blue late-type galaxies'. Fig. \ref{atlas} presents atlas images and spectra of sample galaxies in the 16 fine classes.

\subsection{Sample selection}

\begin{figure}
\includegraphics[width=84mm]{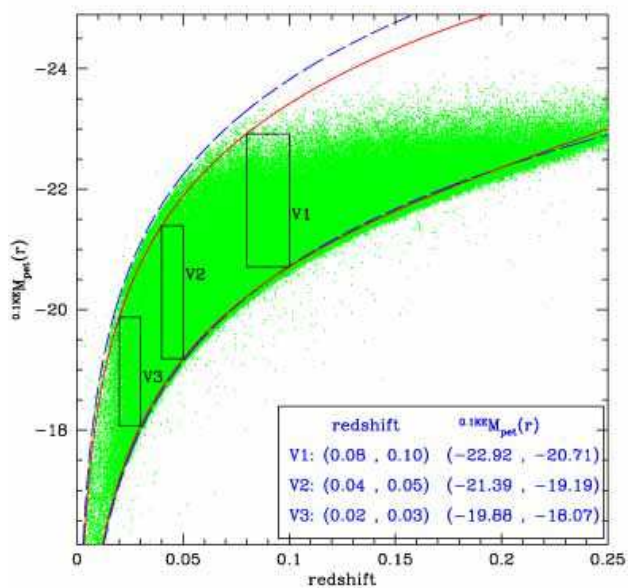}
\caption{Three volumes (V1, V2, and V3) selected in this study. The dashed lines are the limits of the complete spectroscopic sample, and the solid lines are the completeness limits corrected for the redshift effect \citep{bla03} and the galaxy luminosity evolution \citep{teg04}.}
\label{volumes}
\end{figure}

The selection of a galaxy sample is very important, because most properties of galaxies are known to be sensitive to their luminosity and redshift.
We selected three sample volumes in the luminosity versus redshift space as shown in Fig. \ref{volumes}.
Each selected volume is a rectangle because the SDSS spectroscopy has not only a lower-brightness limit but also an upper-brightness limit for completeness.
Each volume has a small redshift range and a large luminosity range, which is adequate to investigate the luminosity dependence of galaxy properties.
Since our sample covers only the nearby universe ($z<0.1$), the effect of the redshift dependence is unlikely to be significant.
Possible small redshift variations of galaxy properties are checked by comparing the difference between the three volumes.
The V1 and V2 volumes in Fig. \ref{volumes} are within the redshift range with reliable spectral information ($0.04<z<0.1$) suggested by \citet{kew06}, but the V3 volume is not, implying that the spectroscopic parameters of galaxies in the V3 volume may be less reliable than those in the V1 or V2 volumes.

\begin{table}
\centering
\caption{The number of galaxies in each class and in each volume}
\label{galnum}
\begin{tabular}{lrrrrr}
\hline \hline
(V1) & REG & BEG & RLG & BLG & Total\\
\hline
Passive & 5109 & 26 & 446 & 77	&	5658\\
H{\protect\scriptsize II} & 2479 & 148 & 1626 & 9632 &	13885\\
Seyfert & 1477 & 98 & 1770 & 1472 &	4817\\
LINER & 4829 & 159 & 3569 & 1281 &	9838\\
\hline
Total & 13894 & 431 & 7411 & 12462 &	34198 \\
\hline \hline
(V2) & REG & BEG & RLG & BLG & Total\\
\hline
Passive & 925 & 23 & 196 & 54	&	1198\\
H{\protect\scriptsize II} & 705 & 141 & 815 & 6348 &	8009\\
Seyfert & 302 & 65 & 547 & 396 &	1310\\
LINER & 824 & 32 & 846 & 233 &	1935\\
\hline
Total & 2756 & 261 & 2404 & 7031 &	12452 \\
\hline \hline
(V3) & REG & BEG & RLG & BLG & Total \\
\hline
Passive & 116 & 40 & 66 & 45	&	267\\
H{\protect\scriptsize II} & 189 & 253 & 325 & 3997 &	4764\\
Seyfert & 45 & 26 & 78 & 76 &	225\\
LINER & 72 & 14 & 108 & 41 &	235\\
\hline
Total & 422 & 333 & 577 & 4159 &	5491 \\
\hline \hline
\end{tabular}
\end{table}

\section{Optical properties}\label{result}

\subsection{Class composition}\label{Lnum}

Table \ref{galnum} lists the number of galaxies in each class and in each volume. The class composition varies with respect to the volume. The three most dominant classes are $h$BLGs (28.2 per cent), $p$REGs (14.9 per cent) and $l$REGs (14.1 per cent) in V1, while they are $h$BLGs (51.0 per cent), $p$REGs (7.4 per cent) and $l$RLGs (6.8 per cent) in V2, and $h$BLGs (72.8 per cent), $h$RLGs (5.9 per cent) and $h$BEGs (4.6 per cent) in V3. 
%Both luminosity and redshift ranges of each volume can be responsible for the difference in the class composition between the volumes.

\begin{figure}
\includegraphics[width=84mm,height=110mm]{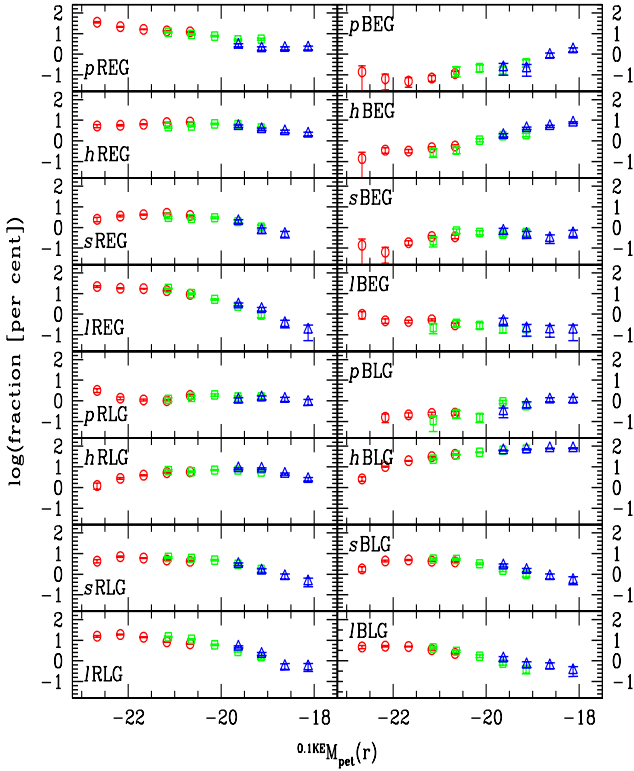}
\caption{Class fraction as a function of luminosity. Percentage variation of each class with respect to luminosity was derived for three volumes: V1 (open circle), V2 (open rectangle), and V3 (open triangle). Each errorbar represents the Poisson error.}
\label{frac1}
\end{figure}

Since most properties of galaxies are known to depend on their luminosity, it is necessary to study the variation of class properties with respect to their luminosity. Fig. \ref{frac1} presents the class fraction versus luminosity in each volume, showing that the variation of the class fraction is continuous even between different volumes. Small discontinuities are found in several classes, but those discontinuities are not significant, considering the fractional uncertainty based on the Poisson error. These results indicate that the effect of luminosity is more important than that of redshift on the difference in the class composition between different volumes.

\citet{cho07} showed that the fraction of early-type galaxies decreases as luminosity decreases, which is consistent with our result.
In Fig. \ref{frac1}, however, we found that such a trend appears to be better established between the colour classes, than between the morphology classes, among passive galaxies. In other words, passive red galaxies are dominant at the bright end, while passive blue galaxies are dominant at the faint end, on average. 
The class fraction distribution of H{\protect\scriptsize II} blue galaxies is similar to that of passive blue galaxies, but the class fraction of H{\protect\scriptsize II} red galaxies is highest at $-20.5 \lesssim \ ^{0.1\textrm{\protect\scriptsize K}}M_{\textrm{\protect\scriptsize pet}}(r) \lesssim -19.5$, unlike that of passive red galaxies.
It is interesting that the fraction of AGN host BLGs shows decrease at the faint end, unlike that of non-AGN BLGs.

\subsection{Luminosity dependence of optical properties}\label{Ldep}

To understand the individual characteristics of each class, we investigate their photometric, structural, and spectroscopic properties, using seven selected physical quantities. Figs. \ref{Lur} -- \ref{Ld4000} show the variation of each quantity with respect to luminosity for the 16 classes, respectively. To reduce the biases in those quantities due to internal extinction in late-type galaxies \citep{cho07}, late-type galaxies with axis ratio smaller than 0.6 are not used in the analysis of each physical quantity, except for axis ratio itself. To reduce the effect of abnormal data points, we use the median statistics.

\subsubsection{Optical colour}\label{SLur}

\begin{figure}
\includegraphics[width=84mm,height=110mm]{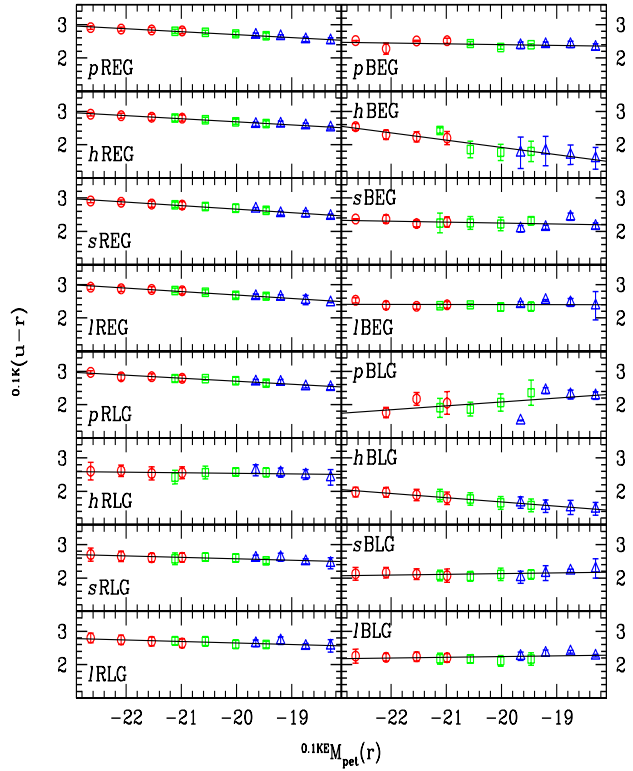}
\caption{$^{0.1\textrm{\protect\scriptsize K}}(u-r)$ colour variation with respect to $^{0.1\textrm{\protect\scriptsize KE}}M_{\textrm{\protect\scriptsize pet}}$, for each class. Each open circle shows the median value at give magnitude bin in V1, and open rectangle and open triangle do in V2 and V3, respectively. Each errorbar represents the sample inter-quartile range (SIQR) of $^{0.1\textrm{\protect\scriptsize K}}(u-r)$ colour at given magnitude bin. The line across symbols in each panel is the linear least-square fit.}
\label{Lur}
\end{figure}

\begin{table}
\centering
\caption{Slopes of the linear fits in the $^{0.1\textrm{\protect\scriptsize K}}(u-r)$ versus $^{0.1\textrm{\protect\scriptsize KE}}M_{\textrm{\protect\scriptsize pet}}$ plots}
\label{Lurfit}
\begin{tabular}{lcc}
\hline \hline
& REG & BEG \\
\hline
Passive & $-0.087\pm0.005(2.792)$	&	$-0.023\pm0.018(2.426)$	\\
H{\protect\scriptsize II} & $-0.092\pm0.004(2.776)$	&	$-0.218\pm0.028(2.139)$	\\
Seyfert & $-0.104\pm0.006(2.771)$	&	$-0.029\pm0.023(2.275)$	\\
LINER & $-0.104\pm0.006(2.792)$	&	$-0.002\pm0.016(2.407)$	\\
\hline \hline
& RLG & BLG \\
\hline
Passive & $-0.092\pm0.007(2.792)$	&	$0.115\pm0.067(1.967)$	\\
H{\protect\scriptsize II} & $-0.020\pm0.014(2.559)$	&	$-0.129\pm0.007(1.810)$	\\
Seyfert & $-0.043\pm0.010(2.617)$	&	$0.022\pm0.016(2.110)$	\\
LINER & $-0.046\pm0.010(2.697)$	&	$0.020\pm0.017(2.222)$	\\
\hline \hline
\end{tabular}
\medskip
\\The slopes of the median $^{0.1\textrm{\protect\scriptsize K}}(u-r)$ with respect to $^{0.1\textrm{\protect\scriptsize KE}}M_{\textrm{\protect\scriptsize pet}}$, and $^{0.1\textrm{\protect\scriptsize K}}(u-r)$ at $^{0.1\textrm{\protect\scriptsize KE}}M_{\textrm{\protect\scriptsize pet}}(r)=-21$ within parentheses.
\end{table}

Fig. \ref{Lur} shows the luminosity dependence of $^{0.1\textrm{\protect\scriptsize K}}(u-r)$ colour for each class.
Nine of 16 classes have a clear trend that the fainter galaxies are bluer. Those variations are steady and continuous even in the transition ranges between volumes, implying that there is little effect from the redshift difference between volumes. $p$BLGs are the only class that are clearly redder when fainter, but the deviation in such a trend is somewhat large.
Table \ref{Lurfit} lists the results of linear least-squares fitting for Fig. \ref{Lur}. We note several interesting features in Fig. \ref{Lur} and Table \ref{Lurfit}.

First, red galaxies (except for $h$RLGs) and H{\protect\scriptsize II} blue galaxies show bluer colour at fainter luminosity. The colour variation in $p$REGs, called often the colour-magnitude relation (CMR), was explained in many previous studies. That is, the CMR of $p$REGs may reflect mainly the difference in the metal abundance with respect to the mass of galaxies \citep{kod97,kau98}. The colour variation in other classes shows somewhat complicated trends, possibly affected by both age and metallicity effects. In star-forming galaxies, the fraction of young stellar population may be an important factor making the colour variation, because it is known that galaxies with small mass are generally young in the local universe \citep{ber05,tre05}.

Second, we found that $p$REGs and $h$REGs have a similar slope in their CMR within 1$\sigma$, and that the CMR slope of AGN host REGs is marginally steeper than that of $p$REGs. The difference in the CMR slope between different spectral classes of REGs is relatively small, compared to those of BEGs, RLGs and BLGs, showing that the metallicity effect may be dominant in REGs, irrespective of their spectral class.
In other words, the effects of star formation or AGN activities may be small in REGs.
The CMR slope in $p$REGs is about $-0.09$, which is significantly smaller than those in $h$BLGs (about $-0.13$). This indicates that the CMR slope of age-effect-dominated galaxies is steeper (larger colour variation) than the CMR slope of metallicity-effect-dominated galaxies, confirming the result of \citet{cho07}.

Third, we found that AGN host galaxies in non-REG classes show very small variation in their colour. Particularly, AGN host blue galaxies do not show significant colour variation with respect to luminosity. Since galaxies with small mass are generally expected to have bluer colours than massive galaxies \citep{kod97,ber05,cho07}, such constant colours of AGN host blue galaxies are somewhat unusual. It is inferred that AGN activity may be responsible for the colour constancy, and that such AGN effects may be most prominent in blue galaxies.

Fourth, $h$BEGs show the largest variation in colour due to the very blue faint members of the class. The very blue faint $h$BEGs were also identified by \citet{cho07}, which may have much younger mean stellar ages or much poorer metal abundance than bright $h$BEGs.

Finally, we found that $p$BEGs and $h$RLGs do not show significant variation in their colour. Since both the metallicity-magnitude relation and the age-magnitude relation are known to cause the slope in CMR, the CMR of $p$BEGs and $h$RLGs may not be the result of any single mechanism. In other words, the combined effects of age and metallicity working differentially with respect to luminosity are one possible origin of the CMR of $p$BEGs and $h$RLGs. Another possibility is differential dust extinction with respect to luminosity.

\subsubsection{Colour gradient}\label{TLdgi}

\begin{figure}
\includegraphics[width=84mm,height=110mm]{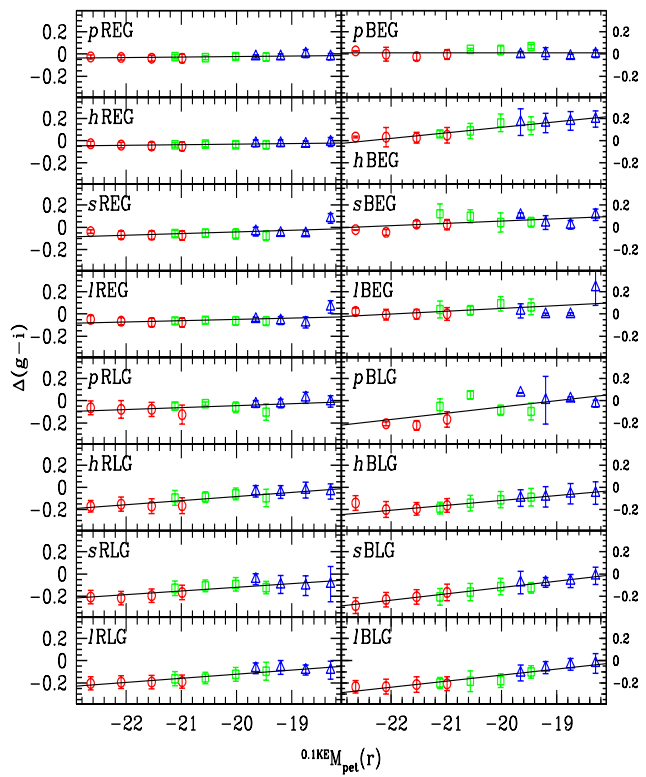}
\caption{The same as Fig. \ref{Lur}, but $\Delta (g-i)$ as Y-axis.}
\label{Ldgi}
\end{figure}

\begin{table}
\centering
\caption{ As Table \ref{Lurfit} but for $\Delta (g-i)$ }
\label{Ldgifit}
\begin{tabular}{lcc}
\hline \hline
& REG & BEG \\
\hline
Passive & $0.004\pm0.002(-0.028)$	&	$0.000\pm0.007(0.011)$	\\
H{\protect\scriptsize II} & $0.005\pm0.002(-0.036)$	&	$0.050\pm0.005(0.071)$	\\
Seyfert & $0.015\pm0.009(-0.056)$	&	$0.020\pm0.011(0.036)$	\\
LINER & $0.012\pm0.008(-0.060)$	&	$0.025\pm0.014(0.026)$	\\
\hline \hline
& RLG & BLG \\
\hline
Passive & $0.018\pm0.008(-0.063)$	&	$0.056\pm0.021(-0.112)$	\\
H{\protect\scriptsize II} & $0.037\pm0.006(-0.118)$	&	$0.044\pm0.002(-0.163)$	\\
Seyfert & $0.033\pm0.007(-0.152)$	&	$0.057\pm0.004(-0.180)$	\\
LINER & $0.037\pm0.005(-0.158)$	&	$0.055\pm0.006(-0.184)$	\\
\hline \hline
\end{tabular}
\end{table}

In Fig. \ref{Ldgi}, the luminosity dependence of $(g-i)$ colour gradient for each class is shown, and the linear fits in Fig. \ref{Ldgi} are summarised in Table \ref{Ldgifit}.
As found by \citet{cho07}, REGs have negative colour gradients (blue outside) on average, and show little variation of their colour gradients with respect to luminosity. Such a colour gradient in REGs is consistent with the previous studies, and may originate from the internal metallicity gradient \citep{tam03,lab05}.
The centres of BEGs are bluer than their outer parts, which results from the shape of the domain in which early-type galaxies were selected (Fig. \ref{class1}). \citet{cho07} reported that the colour gradient of BEGs increases as luminosity decreases, and we additionally found that such a trend is most conspicuous in $h$BEGs. This implies that star formation activity in BEGs may be centrally concentrated, and that the central star formation activity in faint BEGs may be more vigorous than those in bright BEGs.

All sub-classes of late-type galaxies have negative colour gradients (i.e. bluer outside than centre) and show significantly increasing colour gradients as luminosity decreases. The increase in BLGs is slightly larger than that in RLGs. Faint late-type galaxies show very small colour difference between their centre and outside, which implies that the fraction of Scd- and Im-type galaxies in late-type galaxies may increase as luminosity decreases, as pointed out by \citet{cho07}. It is noted that AGN host BLGs show variation in their colour gradient larger than $h$BLGs, which is mainly due to the (negatively) large colour gradients of bright AGN host BLGs. This shows that bright AGN host BLGs may have vigorous star formation activity in their discs, or that the gas cooling in their galactic centre may be suppressed by AGN feedback \citep{cat07}. The colours of AGNs cannot be responsible for those trends in colour gradients, because the colours of AGNs are known to be typically blue \citep{im07}, which would cause the opposite trend.

\subsubsection{Light concentration}

\begin{figure}
\includegraphics[width=84mm,height=110mm]{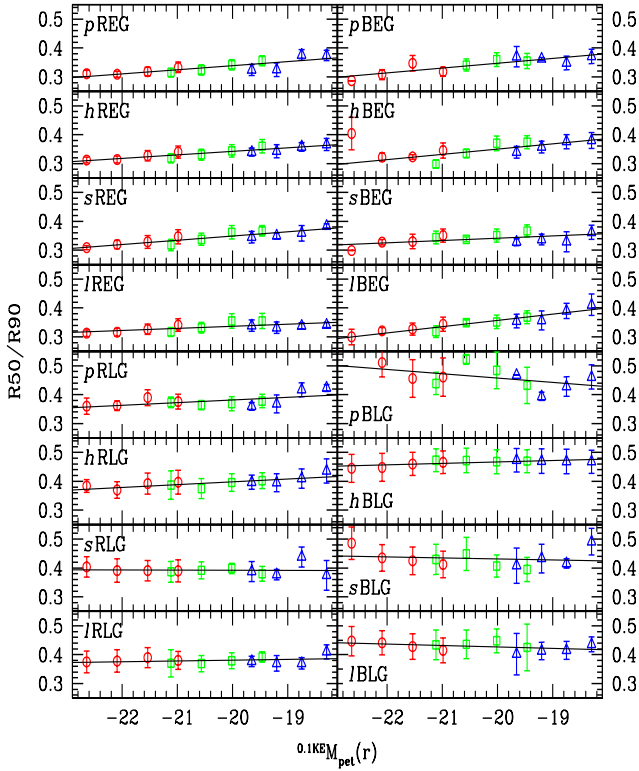}
\caption{The same as Fig. \ref{Lur}, but inverse concentration as Y-axis.}
\label{Lcon}
\end{figure}

\begin{table}
\centering
\caption{ As Table \ref{Lurfit} but for inverse concentration }
\label{Lconfit}
\begin{tabular}{lcc}
\hline \hline
& REG & BEG \\
\hline
Passive & $0.014\pm0.003(0.325)$	&	$0.017\pm0.003(0.331)$	\\
H{\protect\scriptsize II} & $0.012\pm0.002(0.330)$	&	$0.018\pm0.004(0.332)$	\\
Seyfert & $0.015\pm0.002(0.335)$	&	$0.008\pm0.003(0.334)$	\\
LINER & $0.007\pm0.002(0.329)$	&	$0.022\pm0.002(0.335)$	\\
\hline \hline
& RLG & BLG \\
\hline
Passive & $0.009\pm0.004(0.372)$	&	$-0.015\pm0.009(0.472)$	\\
H{\protect\scriptsize II} & $0.010\pm0.002(0.389)$	&	$0.005\pm0.002(0.462)$	\\
Seyfert & $0.000\pm0.004(0.393)$	&	$-0.004\pm0.007(0.435)$	\\
LINER & $0.003\pm0.003(0.378)$	&	$-0.005\pm0.003(0.431)$	\\
\hline \hline
\end{tabular}
\end{table}

The luminosity dependence of inverse concentration for each class is presented in Fig. \ref{Lcon}. Small discontinuities in the variation of inverse concentration are found between different volumes. This may be mainly because we did not conduct \emph{morphological k-correction} in estimating the inverse concentration. Table \ref{Lconfit} lists the linear fits in Fig. \ref{Lcon}.
In early-type galaxies, there is a clear trend that fainter galaxies are less concentrated, as expected from the early-type selection criteria and consistent with the well-known non-homology in early-type galaxies \citep{kor89,mic94}. It is notable that no clear difference is found between $p$REGs, $h$REGs, $p$BEGs and $h$BEGs, which may indicate that the structures of these galaxies are similar. However, this does not necessarily mean that the mass profiles of REGs and BEGs are similar, because BEGs may have young and bright stellar populations with relatively low mass-to-light ratio in their centre.
\citet{cho07} showed that the brightest early-type galaxies are less concentrated, which may be due to recent mergers. We found that such a trend is most conspicuous in $h$BEGs, indicating that the star formation activity in bright $h$BEGs may be related to recent mergers.
It is interesting that faint $l$BEGs are significantly less concentrated than faint $l$REGs, which shows a possibility that faint $l$BEGs may have relatively large disc components.

The concentration of late-type galaxies shows larger scatters than that of early-type galaxies, which may be partially due to the uncertainty in estimating the Petrosian radius of late-type galaxies that have typically double components (bulge+disc) in their surface profile. On average, RLGs are less concentrated than early-type galaxies, and BLGs are the least concentrated.
The fact that RLGs with axis ratio $>0.6$ are more concentrated than BLGs with axis ratio $>0.6$ shows that the bulge fraction may be an important factor determining the colour of a late-type galaxy with small (i.e. close to face-on) inclination.

\subsubsection{Axis ratio}

\begin{figure}
\includegraphics[width=84mm,height=110mm]{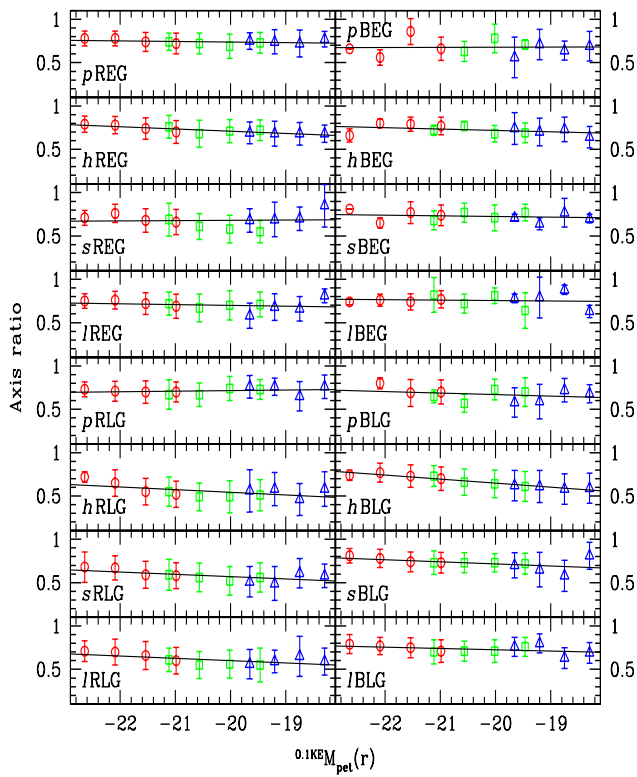}
\caption{The same as Fig. \ref{Lur}, but axis ratio as Y-axis.}
\label{Laxis}
\end{figure}

\begin{table}
\centering
\caption{As Table \ref{Lurfit} but for axis ratio }
\label{Laxisfit}
\begin{tabular}{lcc}
\hline \hline
& REG & BEG \\
\hline
Passive & $-0.007\pm0.006(0.744)$	&	$0.002\pm0.021(0.677)$	\\
H{\protect\scriptsize II} & $-0.025\pm0.005(0.734)$	&	$-0.016\pm0.011(0.734)$	\\
Seyfert & $0.003\pm0.018(0.678)$	&	$-0.008\pm0.012(0.732)$	\\
LINER & $-0.009\pm0.013(0.710)$	&	$-0.004\pm0.017(0.760)$	\\
\hline \hline
& RLG & BLG \\
\hline
Passive & $0.007\pm0.009(0.711)$	&	$-0.017\pm0.019(0.688)$	\\
H{\protect\scriptsize II} & $-0.031\pm0.014(0.575)$	&	$-0.047\pm0.004(0.695)$	\\
Seyfert & $-0.027\pm0.010(0.596)$	&	$-0.024\pm0.013(0.741)$	\\
LINER & $-0.026\pm0.010(0.628)$	&	$-0.015\pm0.010(0.742)$	\\
\hline \hline
\end{tabular}
\end{table}

Fig. \ref{Laxis} shows the luminosity dependence of axis ratio for each class.
In this figure, the axis ratio cut ($>0.6$) was \emph{not} applied to late-type galaxies, unlike the figures for the other parameters. Table \ref{Laxisfit} summarises the linear fits in Fig. \ref{Laxis}.
It is interesting that RLGs have smaller axis ratios than BLGs on average, which may be partially due to the reddening of late-type galaxies with large inclination \citep{bai08}. In other words, late-type galaxies with large inclination would be classified as RLGs, although they had blue colour in the face-on view.
We found that the axis ratio of $h$REGs show significant variation with respect to luminosity, implying that faint $h$REGs may have larger disc components than bright $h$REGs, on average.

In addition to $h$REGs, an obvious trend that fainter galaxies have smaller axis ratio is also found in $h$BLGs. One possibility is that disc components may be more dominant in fainter $h$BLGs, because the axis ratio of galaxies with large inclination and a small bulge may appear smaller than that of galaxies with large inclination and a large bulge.
However, it is also possible that this trend is caused by an inclination effect \citep{cho07}. In other words, since faint $h$BLGs are intrinsically bluer than bright BLGs (see Fig. \ref{Lur}), bright $h$BLGs with large inclination are more easily classified as $h$RLGs than faint $h$BLGs with large inclination, which makes the average axis ratio of faint $h$BLGs smaller than that of bright $h$BLGs.
However, the average axis ratio of faint $h$RLGs is not particularly larger than that of bright $h$RLGs. On the contrary, faint $h$RLGs have marginally smaller axis ratio than bright $h$RLGs. This may be because $h$RLGs with large inclination may suffer dimming due to the extinction, causing the small average axis ratio of faint RLGs. This effect is well discussed in \citet{cho07}.

\subsubsection{Velocity dispersion}

\begin{figure}
\includegraphics[width=84mm,height=110mm]{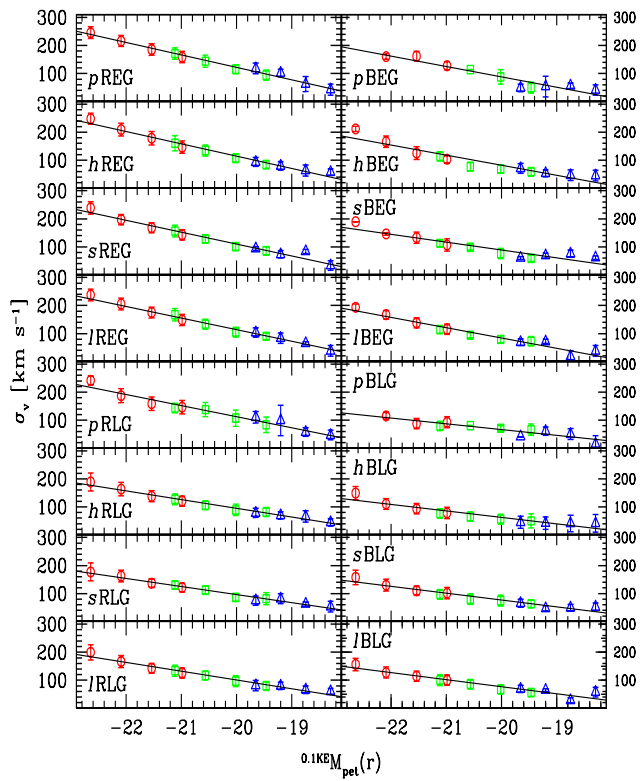}
\caption{The same as Fig. \ref{Lur}, but the velocity dispersion as Y-axis. The velocity dispersion was NOT corrected for the aperture effect.}
\label{Lvd}
\end{figure}

\begin{table}
\centering
\caption{As Table \ref{Lurfit} but for velocity dispersion}
\label{Lvdfit}
\begin{tabular}{lcc}
\hline \hline
& REG & BEG \\
\hline
Passive & $-44.3\pm1.7(166)$	&	$-36.9\pm4.2(125)$	\\
H{\protect\scriptsize II} & $-43.9\pm2.2(158)$	&	$-36.0\pm4.0(118)$	\\
Seyfert & $-42.1\pm2.6(152)$	&	$-27.9\pm3.5(117)$	\\
LINER & $-41.3\pm1.7(155)$	&	$-36.6\pm2.8(121)$	\\
\hline \hline
& RLG & BLG \\
\hline
Passive & $-39.4\pm2.9(152)$	&	$-20.6\pm3.1(87)$	\\
H{\protect\scriptsize II} & $-31.1\pm1.5(126)$	&	$-23.1\pm2.7(86)$	\\
Seyfert & $-28.8\pm1.5(125)$	&	$-24.5\pm2.1(101)$	\\
LINER & $-31.2\pm1.8(132)$	&	$-25.1\pm2.5(102)$	\\
\hline \hline
\end{tabular}
\end{table}

In Fig. \ref{Lvd}, the luminosity dependence of velocity dispersion ($\sigma_v$) for each class is displayed, and Table \ref{Lvdfit} lists the linear fits in Fig. \ref{Lvd}.
According to Fig. \ref{Lvd} and Table \ref{Lvdfit}, early-type galaxies and red galaxies have large variations in their $\sigma_v$ with respect to luminosity, and have large median $\sigma_v$ at $^{0.1\textrm{\protect\scriptsize KE}}M_{\textrm{\protect\scriptsize pet}}(r)=-21$, compared to late-type galaxies and blue galaxies. For example, REGs have very large variation in their $\sigma_v$ and very large median $\sigma_v$ at $^{0.1\textrm{\protect\scriptsize KE}}M_{\textrm{\protect\scriptsize pet}}(r)=-21$, whereas BLGs have very small variation in their $\sigma_v$ and very small median $\sigma_v$ at $^{0.1\textrm{\protect\scriptsize KE}}M_{\textrm{\protect\scriptsize pet}}(r)=-21$.

One main cause for the difference in the variation slope may be the difference in the dynamical mass profile between classes. Since the aperture size of the SDSS spectroscopy is fixed as $3''$, the $\sigma_v$ of a large galaxy reflects the small fraction of its centre, while the $\sigma_v$ of a small galaxy reflects a relatively large fraction of that galaxy. Therefore, if the dynamical mass profiles of galaxies in a class are not concentrated, the $\sigma_v$  -- $^{0.1\textrm{\protect\scriptsize KE}}M_{\textrm{\protect\scriptsize pet}}(r)$ relation slope in that class may be small, because the mass fraction within $3''$ aperture of a large galaxy may be much smaller than that of a small galaxy, in that class. On the other hand, the $\sigma_v$  -- $^{0.1\textrm{\protect\scriptsize KE}}M_{\textrm{\protect\scriptsize pet}}(r)$ relation slope in a class with a highly-concentrated dynamical mass profile, may be relatively large, because a large galaxy in that class may have relatively large fraction of its mass within $3''$ aperture.

In this sense, the large variation in $\sigma_v$ of REGs shows that their dynamical mass profiles may be highly-concentrated, and the outer parts of REGs may not significantly affect the $\sigma_v$ estimation. On the other hand, the small variation in $\sigma_v$ of BLGs indicates that their dynamical mass profiles may not be concentrated and the outer parts of BLGs may affect significantly the $\sigma_v$ estimation.
However, this interpretation is cautiously suggested, because the different luminosity dependence of galaxy mass-to-light ratio between classes may also influence the difference in slope of the $\sigma_v$  -- $^{0.1\textrm{\protect\scriptsize KE}}M_{\textrm{\protect\scriptsize pet}}(r)$ relation.
The rotation of late-type galaxies may also affect the $\sigma_v$ estimation, but late-type galaxies with small axis ratio ($<0.6$) are not used in this analysis. Therefore, the effect of late-type galaxy rotation should be relatively small.

\subsubsection{H{\protect\scriptsize $\alpha$} equivalent width}

\begin{figure}
\includegraphics[width=84mm,height=110mm]{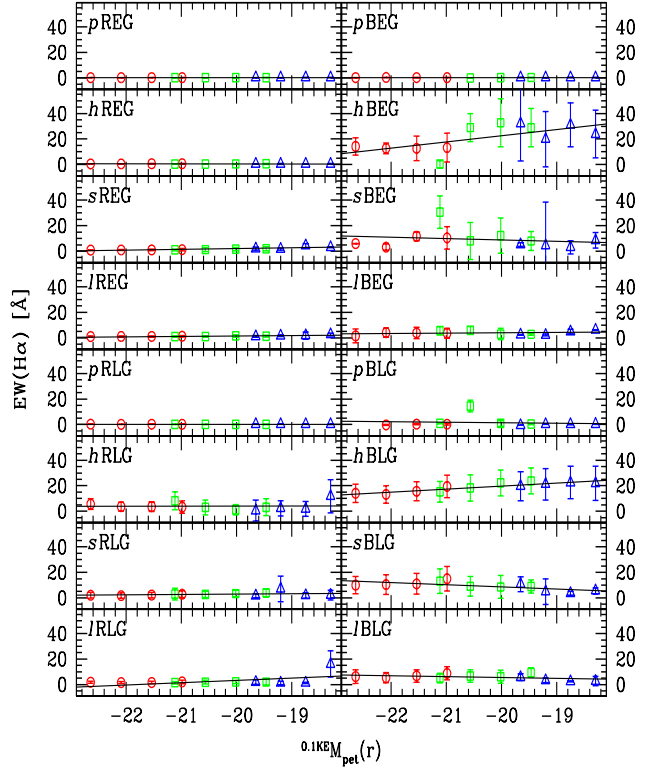}
\caption{The same as Fig. \ref{Lur}, but H{\protect\scriptsize $\alpha$} equivalent width as Y-axis. A positive value of the equivalent width represents an emission line.}
\label{Lha}
\end{figure}

\begin{table}
\centering
\caption{As Table \ref{Lurfit} but for H{\protect\scriptsize $\alpha$} equivalent width }
\label{Lhafit}
\begin{tabular}{lcc}
\hline \hline
& REG & BEG \\
\hline
Passive & $0.01\pm0.01(0.17)$	&	$0.01\pm0.03(0.21)$	\\
H{\protect\scriptsize II} & $0.00\pm0.02(0.49)$	&	$4.88\pm1.85(17.78)$	\\
Seyfert & $0.62\pm0.14(1.37)$	&	$-1.06\pm1.69(9.99)$	\\
LINER & $0.32\pm0.08(1.15)$	&	$0.28\pm0.35(3.67)$	\\
\hline \hline
& RLG & BLG \\
\hline
Passive & $0.00\pm0.01(0.20)$	&	$-0.40\pm1.18(1.88)$	\\
H{\protect\scriptsize II} & $0.07\pm0.74(3.92)$	&	$2.37\pm0.38(17.38)$	\\
Seyfert & $0.33\pm0.33(2.67)$	&	$-1.65\pm0.56(10.30)$	\\
LINER & $1.80\pm1.04(1.46)$	&	$-0.69\pm0.45(6.18)$	\\
\hline \hline
\end{tabular}
\end{table}

The luminosity dependence of H{\protect\scriptsize $\alpha$} equivalent width for each class is presented in Fig. \ref{Lha}, and the linear fits in Fig. \ref{Lha} are summarised in Table \ref{Lhafit}. A positive value of H{\protect\scriptsize $\alpha$} equivalent width indicates line emission, and a negative value indicates line absorption. According to the definition, passive galaxies do not show any significant H{\protect\scriptsize $\alpha$} emission.
We found that red H{\protect\scriptsize II} galaxies (i.e. $h$REGs and $h$RLGs) show small H{\protect\scriptsize $\alpha$} equivalent widths in the almost entire luminosity range, unlike blue H{\protect\scriptsize II} galaxies. The small H{\protect\scriptsize $\alpha$} equivalent width in most red H{\protect\scriptsize II} galaxies implies that those galaxies have just a small fraction of current star formation.

There are trends of increasing H{\protect\scriptsize $\alpha$} equivalent width as luminosity decreases in blue H{\protect\scriptsize II} galaxies, which shows that star formation activity may be more vigorous in faint galaxies, within the spectroscopic fibre aperture. \citet{cho07} suggested that those trends in late-type galaxies may be because the fibre spectra systematically miss the light from the outer discs of bright large galaxies, which is a possible explanation for our $h$BLGs sample. However, the trend in $h$BEGs may be intrinsic rather than caused by the fibre aperture effect, because the star formation activity in $h$BEGs may not be biased to their outer parts, as shown in \S\ref{TLdgi}. The trend of more vigorous star formation activity in fainter galaxies is consistent with previous studies \citep{ber05,tre05}, supporting the galactic downsizing scenario.

The H{\protect\scriptsize $\alpha$} equivalent width of $s$BLGs shows a somewhat unusual trend: it increases from $^{0.1\textrm{\protect\scriptsize KE}}M_{\textrm{\protect\scriptsize pet}}=-23$ to $^{0.1\textrm{\protect\scriptsize KE}}M_{\textrm{\protect\scriptsize pet}}=-21$, but decreases from $^{0.1\textrm{\protect\scriptsize KE}}M_{\textrm{\protect\scriptsize pet}}=-21$ to $^{0.1\textrm{\protect\scriptsize KE}}M_{\textrm{\protect\scriptsize pet}}=-19$.
In fact, the trend in $s$BEGs is similar to that in $s$BLGs, and is therefore possibly related to the luminosity dependence of AGN activity.
However, these trends are not very significant, due to the large error bars.

\subsubsection{4000\AA break}

\begin{figure}
\includegraphics[width=84mm,height=110mm]{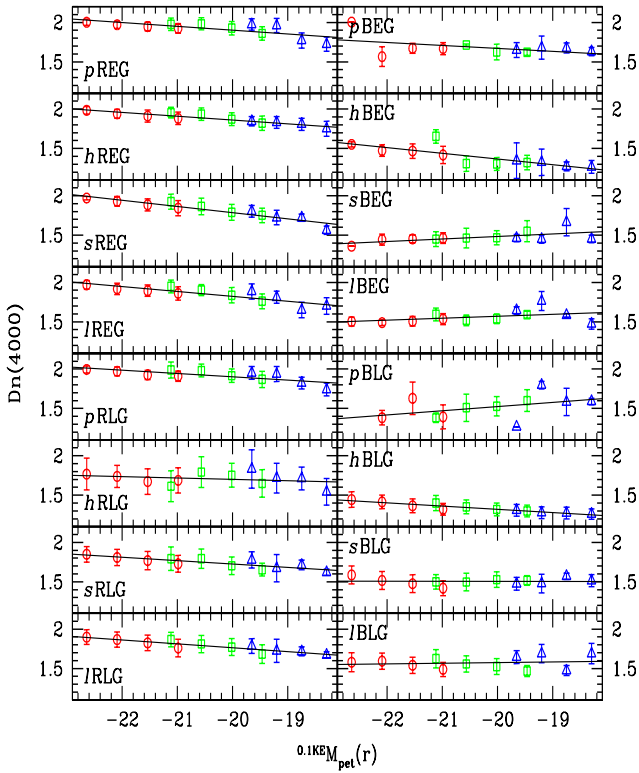}
\caption{The same as Fig. \ref{Lur}, but the 4000{\AA} break index as Y-axis.}
\label{Ld4000}
\end{figure}

\begin{table}
\centering
\caption{As Table \ref{Lurfit} but for 4000{\AA} break index }
\label{Ld4000fit}
\begin{tabular}{lcc}
\hline \hline
& REG & BEG \\
\hline
Passive & $-0.048\pm0.014(1.949)$	&	$-0.038\pm0.024(1.708)$	\\
H{\protect\scriptsize II} & $-0.048\pm0.006(1.909)$	&	$-0.072\pm0.017(1.438)$	\\
Seyfert & $-0.079\pm0.009(1.865)$	&	$0.032\pm0.014(1.452)$	\\
LINER & $-0.063\pm0.011(1.885)$	&	$0.024\pm0.017(1.546)$	\\
\hline \hline
& RLG & BLG \\
\hline
Passive & $-0.042\pm0.011(1.944)$	&	$0.051\pm0.042(1.435)$	\\
H{\protect\scriptsize II} & $-0.018\pm0.018(1.717)$	&	$-0.042\pm0.004(1.363)$	\\
Seyfert & $-0.044\pm0.009(1.769)$	&	$-0.001\pm0.010(1.507)$	\\
LINER & $-0.050\pm0.007(1.814)$	&	$0.008\pm0.018(1.568)$	\\
\hline \hline
\end{tabular}
\end{table}

Fig. \ref{Ld4000} shows the luminosity dependence of the 4000{\AA} break index for each class, and the linear fits in Fig. \ref{Ld4000} are listed in Table \ref{Ld4000fit}.
On average, D$_{n}$(4000) of red galaxies is larger than that of blue galaxies, and D$_{n}$(4000) of passive galaxies is larger than that of non-passive galaxies. Because the D$_{n}$(4000) parameter reflects the fraction of old stellar population in a galaxy, this result shows that red or passive galaxies have larger mean stellar ages than blue or non-passive galaxies.

Faint REGs have smaller D$_{n}$(4000) than bright REGs regardless of their spectral class, which is consistent with galaxy downsizing \citep{cow96,tre05}, in the sense that bright (and maybe massive) galaxies have more old stellar populations than faint (and maybe less-massive) galaxies. This trend is also found in $h$BEGs, $p$RLGs, $s$RLGs, $l$RLGs and $h$BLGs. 
It is noted that blue AGN host galaxies show almost constant D$_{n}$(4000) with respect to luminosity. This may be because the spectral energy distribution (SED) of those blue AGN host galaxies, within the fibre aperture, may be dominated by the AGN SED.
The D$_{n}$(4000) feature in $h$RLGs seems bimodal with large deviation, and the $h$RLGs in the V1 volume show a similar D$_{n}$(4000) variation slope with the other RLGs.

\subsection{Statistics at fixed $\sigma_v$ \label{sigmafix}}

\begin{figure}
\includegraphics[width=84mm]{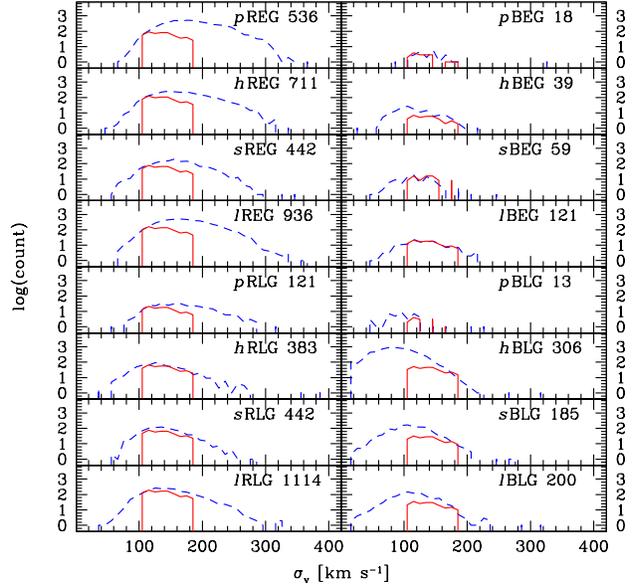}
\caption{Sub-sample selection with the same distribution of velocity dispersion in V1. The dashed line is the distribution of velocity dispersion for each class, and the solid line is the selected sub-sample. In each panel, the class name and sub-sample size are denoted at the upper-right corner.}
\label{VDsam}
\end{figure}

Since most properties of galaxies depend on each other \citep{ber03,cho07}, it is necessary to resample each galaxy class with the same distribution of a \emph{control parameter}, for a direct comparison of galaxy properties between different classes.
One of the most frequently used control parameter is luminosity, which represents basically stellar mass. However, the mass-to-light ratio of a galaxy depends on its bulge-to-disc ratio \citep{yos08}, so that luminosity is not the best as a control parameter in the studies of diverse galaxy classes.
Therefore, we use velocity dispersion instead of luminosity as a control parameter, which represents galaxy dynamical mass.

\begin{figure}
\includegraphics[width=84mm]{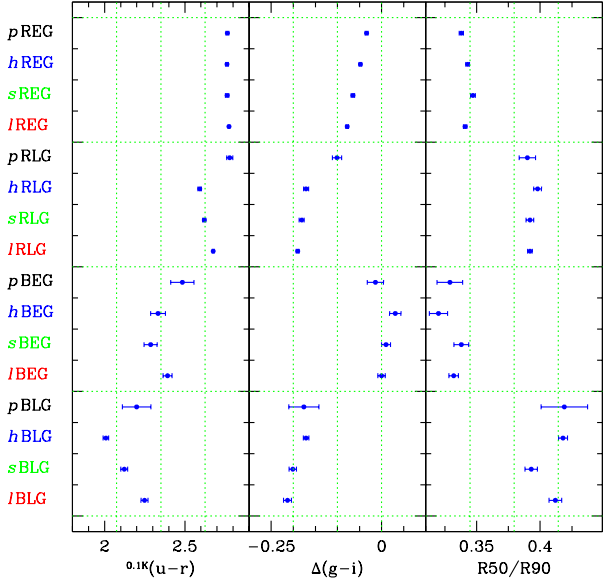}
\caption{Median value and sampling error of three parameters in each class, selected with the same distribution of velocity dispersion as shown in Fig. \ref{VDsam}. The three parameters are $^{0.1\textrm{\protect\scriptsize K}}(u-r)$, $\Delta(g-i)$ and R50/R90. The values are summarised in Table \ref{sumALL}.}
\label{VDstat1}
\end{figure}

\begin{figure}
\includegraphics[width=84mm]{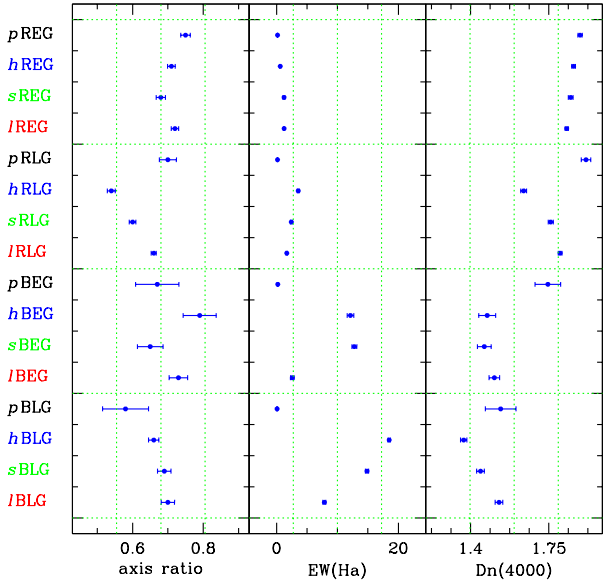}
\caption{The same as Fig. \ref{VDstat1}, but for axis ratio, EW(H{\protect\scriptsize$\alpha$}) [\AA] and D$_{n}$(4000).}
\label{VDstat2}
\end{figure}

Fig. \ref{VDsam} shows the selection of the sub-sample in each class with the same distribution of velocity dispersion. Since the estimation error of $\sigma_{v}$ is very large for $\sigma_{v} < 100$ km s$^{-1}$ \citep{cho07}, we use galaxies with $\sigma_{v} \ge 100$ km s$^{-1}$ only. Since the sample sizes of $p$BEGs, $h$BEGs, $s$BEGs and $p$BLGs are too small, their sub-samples were not selected with the same $\sigma_{v}$ distribution as the sub-samples of the other classes. Therefore, the results in those classes are less reliable than those in the other classes.
In the sub-sample of each class, the median value and the sampling error of six physical quantities were derived as shown in Fig. \ref{VDstat1} and Fig. \ref{VDstat2}.
Each sampling error was estimated by calculating the standard deviation of the median values in 200-times-repetitive sampling.
To reduce the biases in the results due to internal extinction in late-type galaxies \citep{cho07}, late-type galaxies with axis ratio smaller than 0.6 were not used in the analysis of each quantity, except for the axis ratio itself.

We note one possible selection bias. The velocity dispersion of each galaxy may not represent the dynamical mass perfectly, because the velocity dispersions are derived within the limited fibre aperture, and the dynamical mass profiles of galaxies in each class may not be homogeneous. Therefore, the following results may include more or less biases due to the difference in the genuine dynamical mass range of galaxies. Nevertheless, these comparisons are useful, since the velocity dispersion may at least represent the central dynamical mass of each galaxy.

A few quantities should be cautiously compared between different classes, considering the selection criteria of the classes. For example, since optical colour is one criterion to classify our sample galaxies (red galaxies versus blue galaxies), it is meaningful to compare the optical colour only in the same colour class. Similar considerations are necessary in analysing the colour gradient and light concentration.

\subsubsection{Optical colour}

REGs have similar colour regardless of their spectral class, which shows that star formation or AGN activity in REGs is not strong.
In a given morphology-colour class, except for REGs, we found that H{\protect\scriptsize II} galaxies are bluer than passive galaxies, and that LINER galaxies are redder than Seyfert galaxies.
The trend of $^{0.1\textrm{\protect\scriptsize K}}(u-r)$, `REG $>$ RLG $>$ BEG $>$ BLG' is natural, considering the classification of morphological types in Fig. \ref{class1}.
It is noted that $p$RLGs have similar colour to that of REGs.

\subsubsection{Colour gradient}\label{sVDdgi}

Whereas REGs have negative colour gradients (i.e. red centres), as shown by \citet{cho07}, BEGs have positive colour gradients (i.e. blue centres), which is consistent with previous studies \citep{men01,lee06}.
It is interesting that $h$REGs have more negative colour gradients (i.e. bluer outside) than $p$REGs, while $h$BEGs have more positive colour gradients than $p$BEGs. This difference implies that the process of star formation in a galaxy is different between REGs and BEGs. In other words, the star formation in an $h$REG may be dominant in the outer parts of the galaxy, which is possibly triggered by gas infall.
On the other hand, $h$BEGs have star formation mainly in their centres.

We found that AGN host REGs have bluer outsides even than $h$REGs, which shows a possibility that excessive gas infalling into REGs may trigger the AGN activity.
No significant difference is found in the colour gradients between most spectral classes of BEGs, due to the large sampling error, except for $h$BEGs, which have significantly bluer centre than $p$BEGs, and marginally bluer centre than $l$BEGs.

$p$RLGs have very small (but still negative) colour gradients, compared to the other RLGs, showing that the disc components in $p$RLGs may be very small or of red colour. AGN host RLGs have larger negative colour gradients than $h$RLGs, which is possibly caused by the suppression of gas cooling in the centre of AGN host RLGs by AGNs.
Similarly, AGN host BLGs have larger negative colour gradients than non-AGN BLGs.

\subsubsection{Light concentration}

Early-type galaxies have similar concentrations, and are more concentrated than late-type galaxies, as found by \citet{cho07}.
It is interesting that $h$BEGs are most concentrated among early-type galaxies, which may be due to the bright young stellar populations in the centre of $h$BEGs.
Meanwhile, $h$RLGs are significantly more concentrated than $h$BLGs, which imply that many $h$RLGs may be bulge-dominated late-type galaxies. It is noted that $s$BLGs are unusually concentrated, which may be partly due to the existence of a bright AGN in their centre.

\subsubsection{Axis ratio}\label{sVDaxis}

\citet{cho07} showed that early-type galaxies have systematically larger axis ratio than late-type galaxies, which is consistent with our results.
The median axis ratio of $p$REGs is about 0.75, and non-passive REGs have smaller axis ratio than $p$REGs. Considering that non-passive REGs have bluer outsides than $p$REGs (see \S\ref{sVDdgi}), the small axis ratio of non-passive REGs may be due to the existence of small disc components.
On the other hand, $h$BEGs have a median axis ratio larger than that of $p$BEGs (but consistent within $1\sigma$ sampling error), implying that the relationship between $p$BEGs and $h$BEGs may be different from that between $p$REGs and $h$REGs.

The axis ratios of non-passive RLGs are significantly smaller than those of non-passive REGs, in a given spectral class, which implies that disc components in non-passive RLGs may be larger than those in non-passive REGs. Interestingly, the axis ratios of AGN host RLGs are relatively larger than those of $h$RLGs, which may be a selection effect that AGNs are less detected in disc galaxies with large inclination.
It is noted that the difference in the axis ratio between $p$REGs and $p$RLGs is marginal, which shows that most $p$RLGs may be almost face-on or bulge-dominated late-type galaxies.
The axis ratios of non-passive BLGs are larger than those of non-passive RLGs, which is because intrinsic BLGs with large inclination may be classified as RLGs due to the strong dust extinction.

\subsubsection{H{\protect\scriptsize $\alpha$} equivalent width}

According to the definition of the spectral classes in \S\ref{sclass}, the H{\protect\scriptsize $\alpha$} equivalent widths of passive galaxies are almost zero.
Blue H{\protect\scriptsize II} galaxies show large median H{\protect\scriptsize $\alpha$} equivalent widths, implying vigorous star formation activity in them.
On the other hand, red H{\protect\scriptsize II} galaxies have small median H{\protect\scriptsize $\alpha$} equivalent widths, which shows that they may have low current star formation.
The H{\protect\scriptsize $\alpha$} equivalent width of blue Seyfert galaxies is much larger than that of blue LINER galaxies, while the difference in H{\protect\scriptsize $\alpha$} equivalent width between red Seyfert galaxies and red LINER galaxies is not very large.

\subsubsection{4000\AA break}

Morphology class hardly affects the 4000{\AA} break index, while there are clear differences in D$_{n}$(4000) between different colour and spectral classes.
Red galaxies have larger D$_{n}$(4000) than blue galaxies, and passive galaxies have larger D$_{n}$(4000) than non-passive galaxies. The class with the smallest D$_{n}$(4000) is $h$BLG.

\section{Discussion}

\begin{table*}
\centering
\begin{minipage}{126mm}
\caption{Summary of the median values for six parameters in each fine class}
\label{sumALL}
\begin{tabular}{lr@{$\pm$}lr@{$\pm$}lr@{$\pm$}lr@{$\pm$}l}
\hline \hline
& \multicolumn{2}{c}{$p$REG} & \multicolumn{2}{c}{$h$REG}  & \multicolumn{2}{c}{$s$REG} & \multicolumn{2}{c}{$l$REG} \\
\hline
$^{0.1\textrm{\protect\scriptsize K}}(u-r)$ & $2.765$ & $0.009(0.08)$ & $2.762$ & $0.008(0.10)$ & $2.763$ & $0.010(0.11)$ & $2.774$ & $0.006(0.10)$ \\
$\Delta (g-i)$ & $-0.034$ & $0.003(0.03)$ & $-0.048$ & $0.003(0.04)$ & $-0.065$ & $0.004(0.04)$ & $-0.078$ & $0.003(0.04)$ \\
$R50/R90$ & $0.338$ & $0.002(0.02)$ & $0.342$ & $0.001(0.02)$ & $0.347$ & $0.002(0.03)$ & $0.341$ & $0.001(0.02)$ \\
Axis ratio & $0.750$ & $0.013(0.12)$ & $0.720$ & $0.011(0.14)$ & $0.690$ & $0.013(0.16)$ & $0.710$ & $0.010(0.13)$ \\
EW(H{\protect\scriptsize $\alpha$}) [\AA] & $0.169$ & $0.011(0.12)$ & $0.603$ & $0.117(0.34)$ & $1.226$ & $0.148(1.02)$ & $1.242$ & $0.092(0.73)$ \\ 
D$_{n}$(4000) & $1.890$ & $0.009(0.07)$ & $1.861$ & $0.008(0.08)$ & $1.848$ & $0.011(0.10)$ & $1.831$ & $0.008(0.09)$ \\
\hline \hline
& \multicolumn{2}{c}{$p$BEG} & \multicolumn{2}{c}{$h$BEG} & \multicolumn{2}{c}{$s$BEG} & \multicolumn{2}{c}{$l$BEG} \\
\hline
$^{0.1\textrm{\protect\scriptsize K}}(u-r)$ & $2.484$ & $0.073(0.06)$ & $2.333$ & $0.046(0.15)$ & $2.287$ & $0.040(0.12)$ & $2.392$ & $0.027(0.10)$ \\
$\Delta (g-i)$ & $-0.014$ & $0.018(0.03)$ & $0.031$ & $0.013(0.03)$ & $0.010$ & $0.010(0.05)$ & $0.000$ & $0.008(0.04)$ \\ 
$R50/R90$ & $0.329$ & $0.010(0.01)$ & $0.320$ & $0.007(0.01)$ & $0.338$ & $0.006(0.02)$ & $0.332$ & $0.004(0.02)$ \\
Axis ratio & $0.670$ & $0.061(0.07)$ & $0.700$ & $0.047(0.09)$ & $0.640$ & $0.036(0.13)$ & $0.740$ & $0.026(0.09)$ \\
EW(H{\protect\scriptsize $\alpha$}) [\AA] & $0.205$ & $0.059(0.14)$ & $12.137$ & $0.540(5.58)$ & $12.778$ & $0.407(6.21)$ & $2.582$ & $0.283(4.34)$ \\
D$_{n}$(4000) & $1.746$ & $0.057(0.10)$ & $1.475$ & $0.038(0.08)$ & $1.462$ & $0.031(0.07)$ & $1.507$ & $0.023(0.06)$ \\
\hline \hline
& \multicolumn{2}{c}{$p$RLG} & \multicolumn{2}{c}{$h$RLG} & \multicolumn{2}{c}{$s$RLG} & \multicolumn{2}{c}{$l$RLG} \\
\hline
$^{0.1\textrm{\protect\scriptsize K}}(u-r)$ & $2.779$ & $0.019(0.09)$ & $2.592$ & $0.011(0.19)$ & $2.621$ & $0.010(0.13)$ & $2.676$ & $0.006(0.13)$ \\
$\Delta (g-i)$ & $-0.101$ & $0.010(0.08)$ & $-0.171$ & $0.006(0.07)$ & $-0.181$ & $0.006(0.06)$ & $-0.190$ & $0.003(0.06)$ \\
$R50/R90$ & $0.390$ & $0.006(0.03)$ & $0.398$ & $0.003(0.04)$ & $0.392$ & $0.003(0.03)$ & $0.392$ & $0.002(0.04)$ \\
Axis ratio & $0.700$ & $0.024(0.12)$ & $0.540$ & $0.011(0.14)$ & $0.600$ & $0.009(0.14)$ & $0.660$ & $0.007(0.17)$ \\
EW(H{\protect\scriptsize $\alpha$}) [\AA] & $0.160$ & $0.020(0.15)$ & $3.565$ & $0.159(5.16)$ & $2.425$ & $0.166(2.53)$ & $1.687$ & $0.097(1.05)$ \\ 
D$_{n}$(4000) & $1.917$ & $0.021(0.06)$ & $1.638$ & $0.013(0.17)$ & $1.759$ & $0.011(0.13)$ & $1.801$ & $0.008(0.10)$ \\
\hline \hline
& \multicolumn{2}{c}{$p$BLG} & \multicolumn{2}{c}{$h$BLG} & \multicolumn{2}{c}{$s$BLG} & \multicolumn{2}{c}{$l$BLG} \\
\hline
$^{0.1\textrm{\protect\scriptsize K}}(u-r)$ & $2.199$ & $0.089(0.35)$ & $2.008$ & $0.017(0.17)$ & $2.123$ & $0.022(0.13)$ & $2.248$ & $0.022(0.14)$ \\
$\Delta (g-i)$ & $-0.176$ & $0.034(0.07)$ & $-0.171$ & $0.006(0.07)$ & $-0.201$ & $0.008(0.09)$ & $-0.213$ & $0.009(0.08)$ \\
$R50/R90$ & $0.419$ & $0.018(0.09)$ & $0.418$ & $0.004(0.05)$ & $0.393$ & $0.005(0.05)$ & $0.411$ & $0.005(0.03)$ \\
Axis ratio & $0.580$ & $0.065(0.14)$ & $0.660$ & $0.014(0.11)$  & $0.690$ & $0.019(0.09)$ & $0.700$ & $0.019(0.12)$ \\
EW(H{\protect\scriptsize $\alpha$}) [\AA] & $0.087$ & $0.065(0.51)$ & $18.494$ & $0.191(9.27)$ & $14.856$ & $0.275(9.38)$ & $7.858$ & $0.233(5.69)$ \\
D$_{n}$(4000) & $1.535$ & $0.069(0.17)$ & $1.370$ & $0.014(0.09)$ & $1.445$ & $0.017(0.10)$ & $1.528$ & $0.017(0.10)$ \\
\hline\hline
\end{tabular}
\medskip
\\Median value $\pm$ sampling error of each parameter in the sub-samples, selected in Fig. \ref{VDsam}. The value in the parentheses is the SIQR of each parameter in each class.
\end{minipage}
\end{table*}

Tables \ref{sumALL} summarises the median values and sampling errors of six physical quantities in the sub-samples selected with the same distribution of velocity dispersion in Fig. \ref{VDsam}, for REGs, BEGs, RLGs, and BLGs, respectively.
In the following subsections, we discuss the nature of galaxies in each fine class, mainly focusing on the results using the $\sigma_v$-fixed sample (\S\ref{sigmafix}).

\subsection{Red early-type galaxies (REGs)}

$p$REGs may be typical elliptical galaxies, which are old, red and passively-evolving. Other (non-passive) REGs have similar properties to $p$REGs, but there are some notable differences between $p$REGs and non-passive REGs.
First of all, $h$REGs have similar colour to $p$REGs, but their colour gradients are significantly different, in the sense that $h$REGs show larger negative colour gradients than $p$REGs.
This indicates that $h$REGs may have blue populations in their outer parts, implying there may have been some gas infall from their outsides.
AGN host REGs have larger negative colour gradients even than $h$REGs, although the optical colour of AGN host REGs is comparable with that of $p$REGs.

We found that the axis ratio of AGN host REGs is significantly smaller than that of $p$REGs, and the axis ratio of $h$REGs is marginally smaller than that of $p$REGs. These structural features imply that the infalling gas may have formed small disc components in the outer parts of non-passive REGs.
$l$REGs are similar to $s$REGs in their H{\protect\scriptsize $\alpha$} equivalent width and D$_{n}$(4000), but $l$REGs are slightly more concentrated, implying that the disc components may be smaller in $l$REGs than in $s$REGs.
%These properties are consistent with the AGN model of \citet{kew06} that the major difference between Seyferts and LINERs may be the gas accretion rate.

\subsection{Blue early-type galaxies (BEGs)}

BEGs are the galaxies in the blue tail of the early-type galaxies in the colour versus colour gradient diagram \citep{par05}.
BEGs have concentration index comparable to that of REGs, except for $h$BEGs, which are marginally more concentrated than $p$REGs.
The axis ratios of BEGs also agree with those of REGs within very large sampling errors. These features show that BEGs are similar to REGs in their morphological structures.
Compared to REGs, however, BEGs have blue colour, positive colour gradient (blue centre), large H{\protect\scriptsize $\alpha$} equivalent width and small D$_{n}$(4000),
indicating that BEGs have a large number of young stars, compared to REGs.
Since BEGs have bluer centre than REGs, it is considered that the young stellar populations in BEGs may be concentrated toward their centre unlike typical disc galaxies, which is consistent with the previous studies about blue spheroidal galaxies \citep{men01,lee06,cho07}.
BEGs are likely to be the early-type galaxies that recently accreted cold gas from a late-type neighbour during a close encounter or merged with less significant gas-rich galaxies \citep{par08}.

$h$BEGs have a larger positive colour gradient (bluer centre) than $p$BEGs, whereas $h$REGs have a larger negative colour gradient (bluer outside) than $p$REGs. In other words, the star formation of $h$BEGs seems to be dominant in their central regions, while the star formation of $h$REGs seems to be dominant in their outer parts.
It is noted that the structural features of $h$BEGs, $p$BEGs and $p$REGs are similar. The major difference between these galaxy classes is that $p$BEGs have bluer centres and a younger stellar population than $p$REGs, and that $h$BEGs have even bluer centres than $p$BEGs.
However, the star formation activity in $h$BEGs will not continue forever, and the young stellar populations in BEGs will become older and redder as time goes on, if they do not suffer any interactions with other objects.
This implies that $h$BEGs will probably evolve into $p$BEGs after their star formation ends, and that $p$BEGs may also evolve into $p$REGs much later. In other words, $h$BEGs, $p$BEGs and $p$REGs may form the evolutionary sequence of $h$BEGs $\rightarrow$ $p$BEGs $\rightarrow$ $p$REGs, as suggested by \citet{lee06,lee07}.

The axis ratio and concentration of $s$BEGs are in agreement with those of $s$REGs, implying that Seyfert early-type galaxies may have outer disc components. However, we need to be cautious when discussing their structural similarity, because the sampling errors in BEGs are very large.
$l$BEGs show noticeable differences from $s$BEGs: $l$BEGs have marginally larger axis ratios and marginally redder centres than $s$BEGs. This relationship between $l$BEGs and $s$BEGs is similar to that between $l$REGs and $s$REGs.

\subsection{Red late-type galaxies (RLGs)}\label{DRLGs}

RLGs are bluer than REGs.
Considering that the colour gradients of RLGs are significantly more negative than those of REGs, we note that what makes RLGs bluer than REGs may be the blue outer parts of RLGs (i.e. blue disc components). The existence of blue disc components in RLGs is supported by the facts that RLGs are less concentrated than early-type galaxies (for RLGs with axis ratio $> 0.6$) and that the axis ratios of RLGs are smaller than those of early-type galaxies.
Non-passive RLGs are significantly bluer than $p$RLGs.
The concentrations of RLGs are similar among all spectral classes.

The D$_{n}$(4000) of $p$RLGs agrees with that of $p$REGs, but the D$_{n}$(4000) of non-passive RLGs is smaller than that of non-passive REGs, and larger than that of non-passive BEGs. This indicates that the age composition of the stellar populations in RLGs may be intermediate between REGs and BEGs, except for passive galaxies.
We remind the reader that these spectral features are based on the central stellar populations in each galaxy, not entire populations. The D$_{n}$(4000) features of RLGs, therefore, imply that RLGs may have a larger central bulge with old stellar population, compared to BLGs. The fact that $h$RLGs are more concentrated than $h$BLGs also supports this interpretation.

The median axis ratio of RLGs (including RLGs with axis ratio $< 0.6$) is smallest among all of the colour-morphology classes, indicating that there may be many disc galaxies with large inclination within the RLGs. In other words, intrinsic BLGs with large inclination may be classified as RLGs, due to their strong dust extinction.
It is noted that $h$RLGs have axis ratios decreasing as luminosity decreases (see Fig. \ref{Laxis}). Intrinsically bright late-type galaxies with large inclination are dimmed due to the extinction in their discs, causing the small median axis ratio at the faint end.

$p$RLGs have significantly larger axis ratio than $h$RLGs. A good explanation for this difference is that RLGs with large inclination may be classified as $h$RLGs rather than $p$RLGs, because their disc components may be observed within the spectroscopic fibre aperture. That is, many $p$RLGs may be bulge-dominated late-type galaxies with small inclination, so that only their bulge parts were covered in the SDSS spectroscopy. However, $p$RLGs may also include genuinely passive spiral galaxies \citep{cou98, yam04, cho07}.

AGN host RLGs have significantly larger axis ratios than $h$RLGs (but smaller than $p$RLGs). This may be a selection effect caused by AGN obscuration. Since the AGN emission may be difficult to observe in disc galaxies with large inclination, the median axis ratio of the observed AGN host galaxies may be overestimated. In other words, AGN host RLGs with small axis ratio (i.e. large inclination) may be classified as $h$RLGs due to AGN obscuration.

\subsection{Blue late-type galaxies (BLGs)}

Among all colour-morphology classes, BLGs have the bluest colour, the least concentrated profile, the largest H{\protect\scriptsize $\alpha$} equivalent width and the smallest D$_{n}$(4000).
The colour gradient of BLGs is more negative than that of RLGs, except for H{\protect\scriptsize II} galaxies.
Except for passive galaxies, BLGs have larger axis ratio than RLGs, which may be because intrinsic BLGs with large inclination can be classified as RLGs.

$h$BLGs may be typical late-type galaxies with a large amount of current star formation. Their blue colour (particularly in their outer parts), diffuse structure, and large H{\protect\scriptsize $\alpha$} equivalent width are related to the vigorous star formation in the disc of $h$BLGs.
It is difficult to find any significantly different features between $p$BLGs and $h$BLGs, due to the large sampling error of $p$BLGs, except for the H{\protect\scriptsize $\alpha$} equivalent width and D$_{n}$(4000).
Those spectral features show that $p$BLGs may be older than $h$BLGs in their mean stellar age. Some $p$BLGs may be genuinely passive spiral galaxies.

$s$BLGs are significantly more concentrated than $h$BLGs, while the concentration of $l$BLGs is comparable with that of $h$BLGs. This is possibly due to the difference in the brightness of the central AGN between $s$BLGs and $l$BLGs.
The axis ratio of AGN host BLGs is similar to that of AGN host REGs. However, the bulge-to-disc ratio of AGN host BLGs may be smaller than those of AGN host REGs, because the light profile of AGN host BLGs is much less concentrated than those of AGN host REGs. The large median axis ratio of AGN host BLGs may be due to the selection effect caused by AGN obscuration, as mentioned in \S\ref{DRLGs}.

\section{Conclusions}

We conducted a comprehensive study of the nature of the SDSS galaxies in various classes based on their morphology, colour and spectral features.
Using three criteria, we classified the SDSS galaxies into early-type and late-type; red and blue; passive, H{\protect\scriptsize II}, Seyfert and LINER, resulting in 16 fine classes of galaxies in total.
We estimated the luminosity dependence of seven physical quantities in each class, and compared the properties among classes, using a sub-sample with the same distribution of the velocity dispersion.
From the analysis, we found that each galaxy class has its own distinguishable features. This shows that an analysis based on a simple classification may have a risk of mixing up different kinds of objects with different natures.

The red early-type galaxies include well-known typical elliptical galaxies ($p$REGs), but some REGs show evidence for additional star formation in their outer regions ($h$REGs). Some other REGs with AGNs ($s$REGs, $l$REGs) have structural properties showing the existence of larger disc components than $h$REGs, indicating the relationship between AGN activity and gas accretion.
The blue early-type galaxies may be in the process of bulge formation.
The structural similarity between $p$REGs, $p$BEGs and $h$BEGs, supports an evolutionary sequence of $h$BEGs $\rightarrow$ $p$BEGs $\rightarrow$ $p$REGs.
Seyfert early-type galaxies have a close relationship with the outer disc components of early-type galaxies, and the disc components in LINER early-type galaxies ($l$REGs, $l$BEGs) are smaller than those in Seyfert early-type galaxies, on average.
The blue late-type galaxies have properties in agreement with typical spiral galaxies. Most of them are star-forming ($h$BLGs), but a very small fraction of BLGs do not show any evidence of current star formation ($p$BLGs). Some of BLGs have an AGN ($s$BLGs, $l$BLGs), which are less detected at large inclination, and $s$BLGs show particularly bright centres on average.
The median axis ratio in each class shows that some intrinsic BLGs with large inclination may often be classified as red late-type galaxies, due to strong extinction by dust in their disc. In addition to dust extinction, a large bulge-to-disc ratio may make a late-type galaxy red. Many $p$RLGs seem to be bulge-dominated late-type galaxies with small inclination, in which line emission is not detected due to the limited size of the spectroscopic fibre aperture.
Like in BLGs, AGN activity is detected in some RLGs ($s$RLGs, $l$RLGs), which have small inclination.

This paper is the first in the series of comprehensive studies on the nature of the SDSS galaxies in finely-divided classes.
In the following papers, we will inspect various aspects of galaxies in these classes, focusing on their multi-wavelength properties and environmental effects.

\section*{Acknowledgements}

This work was supported in part by a grant (R01-2007-000-20336-0) from the Basic Research Program of the Korea Science and Engineering Foundation (KOSEF).
CBP and YYC acknowledge the support of the KOSEF through the Astrophysical Research Centre for the Structure and Evolution of the Cosmos (ARCSEC).
Funding for the SDSS and SDSS-II has been provided by the Alfred P. Sloan Foundation, 
the Participating Institutions, the National Science Foundation, 
the US Department of Energy, the National Aeronautics and Space Administration, 
the Japanese Monbukagakusho, the Max Planck Society, and the Higher Education Funding Council for England.
The SDSS Web site is http://www.sdss.org/.
The SDSS is managed by the Astrophysical Research Consortium for the Participating Institutions. 
The Participating Institutions are the American Museum of Natural History, 
Astrophysical Institute Potsdam, the University of Basel, the University of Cambridge, 
Case Western Reserve University, the University of Chicago, Drexel University, Fermilab, 
the Institute for Advanced Study, the Japan Participation Group, Johns Hopkins University, 
the Joint Institute for Nuclear Astrophysics, the Kavli Institute for Particle Astrophysics and Cosmology, 
the Korean Scientist Group, the Chinese Academy of Sciences (LAMOST), Los Alamos National Laboratory, 
the Max-Planck-Institute for Astronomy (MPIA), the Max Planck Institute for Astrophysics (MPA), 
New Mexico State University, Ohio State University, the University of Pittsburgh, the University of Portsmouth, 
Princeton University, the US Naval Observatory, and the University of Washington.

\label{lastpage}

\end{document}